\documentclass[a4paper,12pt]{article}

\input{paper_preamble.sty}
\input{short_cut.sty}
\input{aastexv6.sty}

\usepackage{cancel}

\title{Affine connections for Galilean and Carrollian structures: a unified perspective}

\author[1,2]{{Quentin} {Vigneron}\footnote{\href{mailto:quentin.vigneron@umk.pl}{quentin.vigneron@umk.pl}}}
\author[2]{{Hamed} {Barzegar}\footnote{\href{mailto:hamed.barzegar@ens-lyon.fr}{hamed.barzegar@ens-lyon.fr}}}
\author[3]{{James} {Read}\footnote{\href{mailto:james.read@philosophy.ox.ac.uk}{james.read@philosophy.ox.ac.uk}}}

\affil[1]{\small\it{Institute of Astronomy, Faculty of Physics, Astronomy and Informatics}, {Nicolaus Copernicus University}, {{Grudzi{\k{a}}dzka 5}, {Toru\'n}, {87-100} {Poland}}}
\affil[2]{\small\it{ENS de Lyon, CRAL UMR5574, Universit\'e Claude Bernard Lyon 1, CNRS, Lyon, F-69007, France}}
\affil[3]{\small\it{Faculty of Philosophy, University of Oxford\\
Oxford OX2 6GG, United Kingdom}}

\date{}

\begin{document}

\maketitle

\vspace{-1.2cm}
\begin{abstract}
We develop a classification of general Carrollian structures, permitting affine connections with both torsion and non-metricity. We compare with a recent classification of general Galilean structures in order to present a unified perspective on both. Moreover, we demonstrate how both sets of structures emerge from the most general possible Lorentzian structures in their respective limits, and we highlight the role of global hyperbolicity in constraining both structures.  We then leverage this work in order to construct for the first time an ultra-relativistic geometric trinity of gravitational theories, and consider connections which are simultaneously compatible with Galilean and Carrollian structures. We close by outlining a number of open questions and future prospects.
\end{abstract}
\vspace{-.7cm}

\def\contentsname{\empty}
\tableofcontents
\newpage

\section{Introduction}

As is extremely well-known, general relativity (GR) is a spacetime theory formulated using the Levi-Civita affine connection, which is the unique torsion-free connection compatible with the given metric $g_{\mu\nu}$. By now, it is also well-known that it is possible to relax the conditions both of torsion-freeness and of metric compatibility; moreover, it is possible to formulate generalisations of relativistic spacetime physics in terms of said affine connections; to do so is to work in the framework of \emph{metric-affine gravity} (on which see e.g.\ \cite{1995_Hehl_et_al} for a comprehensive review).

In parallel with this axis of generalisation of GR, physicists have become increasingly interested in recent decades in taking various limits of the geometrical structures of GR---in particular, either taking the Galilean ($c\rightarrow \infty$) limit or the Carrollian ($c \rightarrow 0$) limit, to arrive at the structures of Galilean and Carrollian spacetimes, respectively. (For a recent comprehensive and unified study of work on such limits, see \cite{2021_Hansen_PhD}.)
As such, when confronted with both of these threads, it is natural to seek to bring them together, and to consider the most general possible Galilean or Carrollian connections, once the conditions both of torsion-freeness and of non-metricity are relaxed. In the Galilean case, physicists and mathematicians have had control over \emph{torsionful} Galilean connections for several years (see e.g.\ \cite{2016_Bekaert_et_al}), but the most general perspective on Galilean connections, once the metricity condition is also dropped, was offered only very recently by Schwartz \cite{2025_Schwartz}. In the case of Carroll spacetimes (on which see \cite{MarchRead} for a primer), while again the torsionful case has been studied (see \cite{2015_Hartong, 2018_Bekaert_et_al}), as yet there does not exist in the literature any presentation of the most general possible Carrollian structures, permitting both torsion and non-metricity.

The first aim of this article is to fill that gap: we will present a general classification of Carrollian structures permitting affine connections with both torsion and non-metricity, and having done so will compare with the Galilean case as presented in \cite{2025_Schwartz}. The second aim of this article is to show how these general structures emerge from the respective limits of the most general Lorentzian structures.

Although one motivation for this work is simply maximal generality and the exploration of logical space, there are several very direct physics payoffs. Perhaps most notably, one topic of significant discussion in recent years is the `geometric trinity' of gravitational theories, in which one trades the curvature degrees of freedom of GR for either torsion (as in the `teleparallel equivalent of GR' (TEGR)) or non-metricity (as in the `symmetric teleparallel equivalent of GR' (STEGR))---see \cite{2019_Jimenez_et_al} for a review of this topic. Concurrent with the development of the theory of general Galilean connections in \cite{2025_Schwartz}, there was developed in \cite{2024_WRV} a `non-relativistic geometric trinity', in which the curvature degrees of freedom in Newton--Cartan theory (NC) are again traded for either torsion (as in the `teleparallel equivalent of NC' (TENC); see on this also \cite{2023_Schwartz, 2018_Read_et_al}) or non-metricity (as in the `symmetric teleparallel equivalent of NC' (STENC)). Having to hand the full theory of general Carrollian connections opens the possibility of constructing for the first time an \emph{ultra}-relativistic geometric trinity; in this article, we will also complete this task, and show that said trinity is indeed the ultra-relativistic limit of the relativistic geometric trinity (just as for the non-relativistic trinity presented in \cite{2024_WRV}). A second payoff is the ability to consider connections which are compatible with both Galilean and Carrollian structures, and what the physical significance of those connections might be.

Bringing all this together, our plan for the article is as follows. In Section \ref{sec:affineconnections}, we present a general classification of both Galilean and Carrollian structures. In Section \ref{sec_Limits}, we discuss these structures as the limits (respectively, $c\rightarrow \infty$ and $c\rightarrow 0$) of Lorentzian spacetime structures. In Section \ref{sec:trinity}, we show how this work facilitates the construction for the first time of an ultra-relativistic geometric trinity of gravitational theories. In Section \ref{sec:com-connections}, we discuss connections which are simultaneously compatible with Galilean and Carrollian structures. In Section \ref{sec:close}, we wrap up with a number of open questions and future prospects.

\subsection*{Notation}\label{sec:notation}

In this article, we will be working with several different connections: curved, torsionful, and non-metric at the relativistic level, but also at the ultra-relativistic (or Carrollian) level, as well as (to a somewhat lesser extent) at the non-relativistic (or Galilean) level. In order to avoid multiplying notation, we will present our results using the following conventions:
\begin{itemize}
    \item We leave unadorned any affine connection $\Gamma^\mu{}_{\alpha\beta}$ and associated geometric objects which are not \emph{specifically} non-relativistic or ultra-relativistic.
    \item Objects which depend upon non-relativistic (i.e.,\ Galilean) structures (i.e.\ $\tau_\mu$ or $h^{\mu\nu}$) will be denoted with a hat---e.g.,\ $\hat{Q}\indices{_{\mu\nu}} = \nabla_\mu \tau_\nu$ for one of the non-relativistic non-metricities.
    \item  Objects which depend upon ultra-relativistic (i.e.,\ Carrollian) structures (i.e.\ $v^\mu$ or $\gamma_{\mu\nu}$) will be denoted with a check---e.g.,\ $\check{Q}\indices{_{\mu}^{\nu}} = \nabla_\mu v^\nu$ for one of the ultra-relativistic non-metricities.
\end{itemize}

There is a subtlety here which it is worth being completely explicit about. For a \emph{generic} affine connection, there are decompositions available in terms of relativistic contorsion and distorsion tensors, but also in terms of non-relativistic contorsion and distorsion tensors (see \cite[Theorem 4]{2025_Schwartz}), and likewise in terms of ultra-relativistic contorsion and distorsion tensors (see below). These decompositions, to be clear, are available for the {\em very same connection!} As such, it does not make sense to say that a connection \emph{itself} is specifically `non-relativistic' or `ultra-relativistic'; this, indeed, is part of our motivation for adopting the notational conventions above.

When considering the Galilean or Carrollian limits, the relativistic connections ${\Gamma}^\mu{}_{\alpha\beta}$ (and related variables) will be written as Taylor series of the speed of light. The full series will be denoted, respectively, by an upper ``$\epsilon$'' or an upper ``$\lambda$'', e.g., $\accentset{\epsilon}{\Gamma}^\mu{}_{\alpha\beta}$, and the $n$-th order will be denoted by an upper ``$(n)$'', e.g., $\tayl{n}{\Gamma}^\mu{}_{\alpha\beta}$.

Boldface letters are used to represent a tensor in a coordinate-free notation. Throughout this work, we let $\CM$  be a smooth $4$-dimensional manifold.

\section{Affine connections, Galilean and Carrollian structures} \label{sec:affineconnections}

In this section, we present the decomposition of a general affine connection into torsion and non-metricities with respect to a Carrollian structure, and compare it with the Galilean case derived by \citet{2025_Schwartz} (see Table~\ref{tab_Class}). We also introduce the notions of {\it reduced torsion} and {\it reduced non-metricities} allowing us to define contorsion and distorsion (see Table~\ref{tab_Class_reduced} below).

\subsection{Definitions}
 
A {\it Carrollian structure} \citep{1965_Levy_Leblond} is a set $(v^\mu, \gamma_{\mu\nu})$ such that $\T v$ is a nowhere vanishing vector field, $\T\gamma$ is positive semidefinite with $\dim \left(\ker \T \gamma\right) = 1$ and $\T v \in \ker \T \gamma$. We define the {\it  expansion tensor}\footnote{In the literature on Carrollian structures, the intrinsic curvature $K_{\mu\nu} \coloneqq -\Theta_{\mu\nu}$ is more often considered. We choose to use the expansion tensor to avoid any ambiguity with the contorsion tensor $K^\alpha{}_{\mu\nu}$ defined later, which is totally independent of $\Lie{\T v}\gamma_{\mu\nu}$.} of a Carrollian structure as
\begin{align}\label{eq:theta}
    \Theta_{\mu\nu} \coloneqq \frac{1}{2} \Lie{\T v}\gamma_{\mu\nu}\,.
\end{align}

A {\it Galilean structure} \citep{1923_Cartan} is a set $(\tau_\mu, h^{\mu\nu})$ such that $\T \tau$ is a nowhere vanishing 1-form,  $\T h$ is positive semidefinite with $\dim \left(\ker \T h\right) = 1$ and $\T \tau \in \ker \T h$. We define the following object for a Galilean structure
\begin{align}\label{eq:omega-nonmet}
    \omega_{\mu\nu} \coloneqq \partial_{[\mu} \tau_{\nu]}\,.
\end{align}
(In the case of a metric non-relativistic connection, this is of course the temporal torsion. One has to be careful in the non-metric case, however, since torsion and non-metricity are not independent---see below---and as such this object is also associated with non-metricity.)

Given a Carrollian structure $(v^\mu, \gamma_{\mu\nu})$, one can define a dual Galilean structure $(\tau_\mu, h^{\mu\nu})$ (not invariant under Carrollian transformations) such that
\begin{equation}
    \tau_\mu v^\mu = 1 \,,
\quad 
    \tau_\nu h^{\mu\nu} = 0 \,,
\quad 
    h^{\mu\sigma} \gamma_{\sigma\nu} = \delta^\mu_\nu - v^\mu \tau_\nu\,.\label{eq_dual_structure}
\end{equation}
The existence of a 1-form $\T \tau$ with $\tau_\mu v^\mu = 1$ is ensured by the non-vanishing of $\T v$. To show this, it is sufficient to consider a Riemannian metric on $\CM$ whose existence is ensured by the smoothness of $\CM$ (in fact, paracompactness is already sufficient, see e.g.\ \cite[Appendix A]{1984_Wald}). From $\T v(x)\not=0 \ \forall x\in\CM$, the dual 1-form to $\T v$ constructed via the metric is never vanishing and can be normalized so that $\tau_\mu v^\mu = 1$. 
A choice of $\T \tau$ given a Carrollian structure is often referred to as a choice of Ehresmann connection with respect to the Carrollian structure (cf.,\ e.g.,\ \cite{2019_Ciambelli_et_al}).

Similarly, given a Galilean structure $(\tau_\mu, h^{\mu\nu})$, one can define a dual Carrollian structure $(v^\mu, \gamma_{\mu\nu})$ (not invariant under Galilean transformations) such that~\eqref{eq_dual_structure} also hold. Again, the existence of a vector field $\T v$ such that $\tau_\mu v^\mu = 1$ is ensured by the non-vanishing of $\T \tau$. A choice of $\T v$ is often referred to as a choice of timelike observer with respect to the Galilean structure.

Neither structure defines a duality between vectors and 1-forms, i.e.\ it does not define an invertible way of raising and lowering indices. However, as is commonly done in the literature on Carrollian and Galilean structures, we introduce an important convention here: after having defined a new tensor, raised and lowered indices with respect to that definition will always be defined with respect to $h^{\mu\nu}$ and $\gamma_{\mu\nu}$, in Galilean and Carrollian cases, respectively; e.g. $\Theta_\mu{}^\nu \coloneqq \Theta_{\mu\sigma} h^{\sigma\nu}$.

Finally, given two structures $(v^\mu, \gamma_{\mu\nu})$ and $(\tau_\mu, h^{\mu\nu})$, we define the connection
\begin{align}
    {^{v,\tau}}\Gamma^\alpha_{\mu\nu} 
    \coloneqq 
    h^{\alpha\sigma} \left[\partial_{(\mu} \gamma_{\nu)\sigma} - \tfrac{1}{2} \partial_{\sigma} \gamma_{\mu\nu}\right] 
    + 
    v^\alpha \partial_{(\mu} \tau_{\nu)}\,.
\end{align}
This connection is only compatible with $(v^\mu, \gamma_{\mu\nu})$ if $\Theta_{\mu\nu} = 0$ and is only compatible with $(\tau_\mu, h^{\mu\nu})$ if $\omega_{\mu\nu} = 0$.

\subsection{Decomposition of a general affine connection}

The results of this section are summarized in Table~\ref{tab_Class}.

The characterization of general affine connections with respect to Galilean structures was studied by \citet{2025_Schwartz}. Given an affine connection $\T \nabla$ with torsion $T^\alpha{}_{\mu\nu} \coloneqq 2 \Gamma^\alpha_{[\mu\nu]}$ and a Galilean structure $(\tau_\mu, h^{\mu\nu})$, the non-metricities of $\T \nabla$ are defined as
\begin{align}
    \hat Q_\alpha{}^{\mu\nu} \coloneqq \nabla_\alpha h^{\mu\nu}, \quad \hat Q_{\mu\nu} \coloneqq \nabla_\mu \tau_\nu\,. \label{eq_def_Q}
\end{align}
Contrary to the Lorentzian case, non-metricities and torsion are not independent and we have the following identities
\begin{align}
    \tau_\nu \hat Q_\alpha{}^{\mu\nu} = - \hat Q_\alpha{}^\mu, \quad \tau_\alpha T^\alpha{}_{\mu\nu} = -2\hat Q_{[\mu\nu]} + 2\omega_{\mu\nu}\,. \label{eq_identities_Gal}
\end{align}

Given a choice of observer with a dual structure $(v^\mu, \gamma_{\mu\nu})$, \citet{2025_Schwartz} showed that the difference $\Gamma^\alpha_{\mu\nu} - {^{v,\tau}}\Gamma^\alpha_{\mu\nu}$ can be written solely as function of $\hat Q_\alpha{}^{\mu\nu}$, $\hat Q_{\mu\nu}$, $T^\alpha{}_{\mu\nu}$ and a 2-form $\kappa_{\mu\nu}$ as follows
\begin{align}
    \Gamma^\alpha_{\mu\nu} - {^{v,\tau}}\Gamma^\alpha_{\mu\nu} &=h^\alpha{}_\beta \hat Q_{(\mu\nu)}{}^\beta - \frac{1}{2} \hat Q^\alpha{}_{\mu\nu} - v^\alpha \hat Q_{(\mu\nu)} - T_{(\mu\nu)}{}^\alpha + \frac{1}{2} T^\alpha{}_{\mu\nu} + 2\tau_{(\mu} \kappa_{\nu)\beta}h^{\alpha\beta}\, . \label{eq_Classification_Gal}
\end{align}
$\kappa_{\mu\nu} \coloneqq \nabla_{[\mu} v^\alpha\, \gamma_{\nu]\alpha}$ is called the {\it Coriolis field} and encodes the 4-acceleration and the spatial rotation (with respect to the connection $\T \nabla$) of the set of observers described by the 4-velocity $v^\mu$.

In what follows we derive the dual formulae for a Carrollian structure. Given an affine connection $\T \nabla$ with torsion $T^\alpha{}_{\mu\nu} \coloneqq 2 \Gamma^\alpha_{[\mu\nu]}$ and a Carrollian structure $(v^\mu, \gamma_{\mu\nu})$, we define the non-metricities of $\T \nabla$ as
\begin{equation}
    \check Q_{\mu}{}^{\nu} \coloneqq \nabla_\mu v^\nu \,,
\quad 
    \check Q_{\alpha\mu\nu} \coloneqq \nabla_\alpha \gamma_{\mu\nu}\,,
\end{equation}
which satisfy the following identities
\begin{equation}\label{eq_identities_Car}
    v^\nu \check Q_{\alpha\mu\nu} = - \check Q_{\alpha\mu} \,,
\quad 
    v^\alpha T_{(\mu\nu)\alpha} = \frac{1}{2} v^\alpha \check Q_{\alpha\mu\nu} + \check Q_{(\mu\nu)} - \Theta_{\mu\nu} \,.
\end{equation}
A direct computation gives
\begin{align*}
    \Gamma^\alpha_{\mu\nu} - {^{v,\tau}}\Gamma^\alpha_{\mu\nu}
        &= h^{\alpha\sigma} \left[- \nabla_{(\mu}\gamma_{\nu)\sigma} + \tfrac{1}{2}  \nabla_{\sigma} \gamma_{\mu\nu}\right] -v^\alpha \nabla_{(\mu} \tau_{\nu)} \\
            &\qquad - h^{\alpha\sigma}\left[\Gamma^\kappa_{(\mu\nu)} \gamma_{\kappa\sigma} + \gamma_{\kappa(\mu} \Gamma^\kappa_{\nu)\sigma} - \Gamma^\kappa_{\sigma(\mu} \gamma_{\nu)\kappa}\right] - v^\alpha \Gamma^\kappa_{(\mu\nu)} \tau_\kappa + \Gamma^\alpha_{\mu\nu},
\end{align*}
which leads to the following proposition:
    
\begin{proposition} Given a Carrollian structure $(v^\mu, \gamma_{\mu\nu})$ and a choice of dual structure $(\tau_\mu, h^{\mu\nu})$, a general affine connection $\Gamma^\alpha_{\mu\nu}$ can be written as follows:
\begin{equation}
    \begin{aligned}
    \Gamma^\alpha_{\mu\nu} - {^{v,\tau}}\Gamma^\alpha_{\mu\nu}
        &= h^{\alpha\sigma} \left[-\check Q_{(\mu\nu)\sigma} + \tfrac{1}{2} \check Q_{\sigma\mu\nu}\right] - \gamma_{\sigma(\mu}T^\sigma{}_{\nu)\kappa} h^{\kappa\alpha} + \frac{1}{2} T^\alpha{}_{\mu\nu} - v^\alpha {\chi}_{\mu\nu} \\
        &= -\check Q_{(\mu\nu)}{}^\alpha + \frac{1}{2} \check Q^\alpha{}_{\mu\nu} -T_{(\mu\nu)}{}^\alpha + \frac{1}{2} T^\alpha{}_{\mu\nu} - v^\alpha \chi_{\mu\nu}, \label{eq_Gene}
    \end{aligned}
\end{equation}
with ${\chi}_{\mu\nu} \coloneqq \nabla_{(\mu} \tau_{\nu)}$.
\end{proposition}

The dual to the Coriolis field in the Carrollian case is the tensor $\chi_{\mu\nu}$, which likewise encodes the non-uniqueness of a compatible Carrollian connection. The fact that $\chi_{\mu\nu}$ is symmetric (rather than antisymmetric, as for the Coriolis field of a Galilean connection) makes this object somewhat harder to work with, and presents roadblocks to a straightforward understanding of the circumstances under which one has, for example, a potential-based Carrollian spacetime theory (in analogy with standard potential-based Newtonian gravity); we return to this issue in Section \ref{sec:close}.

\begin{table}[t]
\centering\small
\caption{\small Properties of a general affine connection $\T\nabla$ with respect to a Galilean and a Carrollian structure. In both cases, the indices of the non-metricities and of the torsion are raised by $h^{\mu\nu}$ and lowered by $\gamma_{\mu\nu}$.\label{tab_Class}}
\renewcommand{\arraystretch}{1.5}
\newcommand{\vspacetable}[1]{\multicolumn{2}{c}{} \vspace{#1}\\}
\begin{tabular}{p{\columnwidth*1/5-2\tabcolsep}|p{\columnwidth*2/5 - 2\tabcolsep}|p{\columnwidth*2/5- 2\tabcolsep}}
	\hline\hline 
    
\bf \centering Quantities
    &\centering \bf Galilean structure
    & \centering \bf Carrollian structure \arraybackslash\\
\hline\hline

Invariant structures
    & $(\tau_\mu, h^{\mu\nu})$
    & $(v^\mu, \gamma_{\mu\nu})$ \\
\hline 

Gauge dependent structures
    & $(v^\mu, \gamma_{\mu\nu}) \ \Leftrightarrow$ `choice of observer'
    & $(\tau_\mu, h^{\mu\nu}) \ \Leftrightarrow$ `choice of Ehresmann connection'\\
\hline
 
{Intrinsic objects}
    & $\omega_{\mu\nu} = \partial_{[\mu}\tau_{\nu]}$
    & $\Theta_{\mu\nu} = \frac{1}{2} \Lie{\T v}\gamma_{\mu\nu}$ \\
\hline

Non-metricities
    & $\hat Q_{\mu\nu} = \nabla_\mu \tau_\nu \,, \quad \hat Q_\alpha{}^{\mu\nu} = \nabla_\alpha h^{\mu\nu}$
    & $\check Q_{\mu}{}^{\nu} = \nabla_\mu v^\nu \,, \quad \check Q_{\alpha\mu\nu} = \nabla_\alpha \gamma_{\mu\nu}$ \\

\hline

Identities
    & $\begin{aligned}
        &\tau_\nu \hat Q_\alpha{}^{\mu\nu} = - \hat Q_\alpha{}^\mu \\
        &\tau_\alpha T^\alpha{}_{\mu\nu} = -2\hat Q_{[\mu\nu]} + 2\omega_{\mu\nu}
    \end{aligned}$
    & \addstackgap{$\begin{aligned}
        &v^\nu \check Q_{\alpha\mu\nu} = - \check Q_{\alpha\mu} \\
        &v^\alpha T_{(\mu\nu)\alpha} = \tfrac{1}{2} v^\alpha \check Q_{\alpha\mu\nu} + \check Q_{(\mu\nu)} - \Theta_{\mu\nu}
    \end{aligned}$} \\
\hline

$\Gamma^\alpha_{\mu\nu} - {^{v,\tau}}\Gamma^\alpha_{\mu\nu}$
    & $\begin{aligned} 
        &h^\alpha{}_\beta \hat Q_{(\mu\nu)}{}^\beta - \tfrac{1}{2} \hat Q^\alpha{}_{\mu\nu} - v^\alpha \hat Q_{(\mu\nu)} \,, \\ 
        &\quad-T_{(\mu\nu)}{}^\alpha + \tfrac{1}{2} T^\alpha{}_{\mu\nu} + 2\tau_{(\mu} \kappa_{\nu)\beta}h^{\alpha\beta}\\
        &\text{with }\kappa_{\mu\nu} = \nabla_{[\mu} v^\alpha\, \gamma_{\nu]\alpha}.
    \end{aligned}$
    & \addstackgap{$\begin{aligned} 
        &-\check Q_{(\mu\nu)}{}^\alpha + \tfrac{1}{2} \check Q^\alpha{}_{\mu\nu} \\
        &\quad-T_{(\mu\nu)}{}^\alpha + \tfrac{1}{2} T^\alpha{}_{\mu\nu} - v^\alpha \chi_{\mu\nu} \,, \\
        &\text{with }\chi_{\mu\nu} = \nabla_{(\mu} \tau_{\nu)}.
    \end{aligned}$} \\
\hline
\end{tabular}
\end{table}

\subsection{Reduced variables}\label{sec_Reduced}

\begin{table}[t]
\centering\small
\caption{\small Definition of the reduced quantities, contorsion and distorsion.\label{tab_Class_reduced}}
\renewcommand{\arraystretch}{1.5}
\newcommand{\vspacetable}[1]{\multicolumn{2}{c}{} \vspace{#1}\\}
\begin{tabular}{p{\columnwidth*1/5-2\tabcolsep}|p{\columnwidth*17/40-2\tabcolsep}|p{\columnwidth*15/40-2\tabcolsep}}
	\hline\hline 
    
\bf \centering Quantities
    &\centering \bf Galilean structure
    & \centering \bf Carrollian structure \arraybackslash\\
\hline\hline
\multicolumn{3}{c}{\sc Metric case ($\hat Q_{\mu\nu} = 0 = \hat Q_\alpha{}^{\mu\nu}$ {and} $\check Q_{\mu}{}^{\nu} = 0 = \check Q_{\alpha\mu\nu}$)} \vspace{.1cm}\\

Reduced 
    
    torsion
    &
    $\begin{aligned}
        &\hat \CT^\alpha{}_{\mu\nu} = T^\alpha{}_{\mu\nu} - 2v^\alpha  \omega_{\mu\nu}
    \end{aligned}$
    &
    $\begin{aligned}
        &\check \CT^\alpha{}_{\mu\nu} = T^\alpha{}_{\mu\nu} - 2\tau_{[\mu} \Theta_{\nu]\sigma} h^{\alpha\sigma}
    \end{aligned}$\\
\hline
    
Reduced 
    
    identities
    & $\begin{aligned}
         \tau_\alpha \hat \CT^\alpha{}_{\mu\nu} = 0
    \end{aligned}$
    & $\begin{aligned}
        v^\alpha \check \CT_{(\mu\nu)\alpha} = 0
    \end{aligned}$\\ \hline
    
Contorsion
    & $\hat K^\alpha{}_{\mu\nu} = -\hat \CT_{(\mu\nu)}{}^\alpha + \frac{1}{2} \hat \CT^\alpha{}_{\mu\nu}$
    & $\check K^\alpha{}_{\mu\nu} = -\check \CT_{(\mu\nu)}{}^\alpha + \frac{1}{2} \check \CT^\alpha{}_{\mu\nu}$
    \\ \hline

$\Gamma^\alpha_{\mu\nu} - {^{v,\tau}}\Gamma^\alpha_{\mu\nu}$
     & $\begin{aligned} 
        &\hat K^\alpha{}_{\mu\nu} + v^\alpha \omega_{\mu\nu} + 2\tau_{(\mu} \kappa_{\nu)\beta}h^{\alpha\beta}\\
        \end{aligned}$
     & $\begin{aligned} 
            & \check K^\alpha{}_{\mu\nu} - \tau_\nu \Theta_{\mu\sigma}h^{\sigma\alpha} - v^\alpha \chi_{\mu\nu}
    \end{aligned}$ \\
\hline
 
\multicolumn{3}{c}{\sc Symmetric case ($T^\alpha{}_{\mu\nu} = 0$)} \vspace{.1cm}\\

Reduced 
    
    non-metricities
    & $\begin{aligned}
        \hat \CQ_{\mu\nu} &= \hat Q_{\mu\nu} - \omega_{\mu\nu} - 2v^\sigma \omega_{\sigma(\mu} \tau_{\nu)} \\
        \hat \CQ_\alpha{}^{\mu\nu} &= \hat Q_\alpha{}^{\mu\nu} - 4v^\lambda v^{(\mu} h^{\nu)\sigma} \omega_{\sigma(\alpha}\tau_{\lambda)}
    \end{aligned}$
    & $\begin{aligned}
        \check \CQ_\mu{}^\nu &= \check Q_\mu{}^\nu - \Theta_\mu{}^\nu\\
        \check \CQ _{\alpha\mu\nu} &= \check Q_{\alpha\mu\nu} + 2\tau_{(\mu} \Theta_{\nu)\alpha}
    \end{aligned}$\\
\hline
    
Reduced 
    
    identities 
    & $\begin{aligned}
         \tau_\nu \hat \CQ_\alpha{}^{\mu\nu} + \hat \CQ_\alpha{}^\mu &= 0 \\
         \hat \CQ_{[\mu\nu]} &= 0
    \end{aligned}$
    & \addstackgap{$\begin{aligned}
        v^\nu \check \CQ_{\alpha\mu\nu} + \check \CQ_{\alpha\mu} &= 0 \\
        \frac{1}{2} v^\alpha \check \CQ_{\alpha\mu\nu} + \check \CQ_{(\mu\nu)} &= 0
    \end{aligned}$}\\
\hline

Distorsion
    & $\hat L^\alpha{}_{\mu\nu} = h^\alpha{}_\beta \hat \CQ_{(\mu\nu)}{}^\beta - \frac{1}{2} \hat \CQ^\alpha{}_{\mu\nu} - v^\alpha \hat \CQ_{(\mu\nu)}$
    & $\check L^\alpha{}_{\mu\nu} \coloneqq -\check \CQ_{(\mu\nu)}{}^\alpha + \frac{1}{2} \check \CQ^\alpha{}_{\mu\nu}$
    \\
\hline

$\Gamma^\alpha_{\mu\nu} - {^{v,\tau}}\Gamma^\alpha_{\mu\nu}$
     & $\begin{aligned} 
        &\hat L^\alpha{}_{\mu\nu} + 2v^\alpha v^\lambda\omega_{\lambda(\mu}\tau_{\nu)} + 2\tau_{(\mu} \kappa_{\nu)\beta}h^{\alpha\beta}\\
        \end{aligned}$
     & \addstackgap{$\begin{aligned} 
            & \check L^\alpha{}_{\mu\nu} - v^\alpha \chi_{\mu\nu}
    \end{aligned}$} \\
\hline
 
\end{tabular}
\end{table}

The results of this section are summarized in Table~\ref{tab_Class_reduced}.

Due to the identities \eqref{eq_identities_Gal} and \eqref{eq_identities_Car}, the non-metricities and torsion are not independent. In particular, because of the second identities in \eqref{eq_identities_Gal} and \eqref{eq_identities_Car}, in the general case where $\omega_{\mu\nu} \not=0$ and $\Theta_{\mu\nu}\not=0$, one cannot choose both a symmetric and compatible connection, i.e. some components of the torsion and non-metricities depend on $\omega_{\mu\nu}$ and $\Theta_{\mu\nu}$. Removing this dependency would enable us to extract the parts of the torsion and non-metricities that are independent of $\omega_{\mu\nu}$ and $\Theta_{\mu\nu}$, obtaining what we call {\it reduced torsion} (denoted $\hat \CT^\alpha{}_{\mu\nu}$ and $\check \CT^\alpha{}_{\mu\nu}$), and {\it reduced non-metricities} (denoted $\hat \CQ_{\mu\nu}$, $\hat\CQ_\alpha{}^{\mu\nu}$, and $\check \CQ_{\mu}{}^{\nu}$, $\check\CQ_{\alpha\mu\nu}$). Two natural conditions can be considered to define these reduced quantities:
\begin{enumerate}
    \item The differences $\hat Q_{\mu\nu} - \hat \CQ_{\mu\nu}$, $\hat Q_\alpha{}^{\mu\nu} - \hat \CQ_\alpha{}^{\mu\nu}$ and $T^\alpha{}_{\mu\nu} - \hat \CT^\alpha{}_{\mu\nu}$ (respectively $\check Q_{\mu}{}^{\nu} - \check \CQ_{\mu}{}^{\nu}$, $\check Q_{\alpha\mu\nu} - \check \CQ_{\alpha\mu\nu}$ and $T^\alpha{}_{\mu\nu} - \check \CT^\alpha{}_{\mu\nu}$) involve only the tensors $\tau_\mu$, $h^{\mu\nu}$, $v^\mu$ and $\gamma_{\mu\nu}$, and depend linearly on $\omega_{\mu\nu}$ (respectively $\Theta_{\mu\nu}$) in the Galilean case (respectively in the Carrollian case).
    \item The terms related to $\omega_{\mu\nu}$ and $\Theta_{\mu\nu}$ should disappear from~\eqref{eq_identities_Gal} and~\eqref{eq_identities_Car}, hence obtaining the {\it reduced identities}
\begin{align}
    \tau_\nu \hat \CQ_\alpha{}^{\mu\nu} &= - \hat \CQ_\alpha{}^\mu\,,
    &\tau_\alpha \hat \CT^\alpha{}_{\mu\nu} &= -2\CQ_{[\mu\nu]}\,,\label{eq_reduced_identities_Gal} \\
    v^\nu \CQ_{\alpha\mu\nu} &= - \CQ_{\alpha\mu}\,,
    &v^\alpha \CT_{(\mu\nu)\alpha} &= \frac{1}{2} v^\alpha \CQ_{\alpha\mu\nu} + \CQ_{(\mu\nu)}\,.\label{eq_reduced_identities_Car}
\end{align}
\end{enumerate} 
As shown in Appendix~\ref{app_reduced_quantities}, these conditions are not sufficient to give a unique definition for the reduced quantities. 
A way of constraining the degrees of freedom is to define the reduced quantities not for a general affine connection, but when assuming either metricity or no torsion, i.e. defining $\hat \CT^\alpha{}_{\mu\nu}$ and $\check \CT^\alpha{}_{\mu\nu}$ only when $\hat Q_{\mu\nu} = 0 = \hat Q_\alpha{}^{\mu\nu}$ and $\check Q_{\mu}{}^{\nu} = 0 = \check Q_{\alpha\mu\nu}$, and defining $\hat \CQ_{\mu\nu}$, $\hat\CQ_\alpha{}^{\mu\nu}$, $\check \CQ_{\mu}{}^{\nu}$, $\check\CQ_{\alpha\mu\nu}$ only when $T^\alpha{}_{\mu\nu} = 0$. In the Carrollian case this leads to a unique way of defining a reduced torsion $\check\CT^\alpha{}_{\mu\nu}$ as shown in Appendix~\ref{app_reduced_quantities_Car}. However, the remaining reduced quantities ($\hat\CT^\alpha{}_{\mu\nu}$, $\hat \CQ_{\mu\nu}$, $\hat\CQ_\alpha{}^{\mu\nu}$, and $\check \CQ_{\mu}{}^{\nu}$, $\check\CQ_{\alpha\mu\nu}$) remain non-uniquely defined. Further constraining the freedom in their definition can be done with the Galilean and Carrollian limits detailed in Section~\ref{sec_Limits}.

The Galilean/Carrollian limit of the Levi-Civita connection of a Lorentzian structure is a torsion-free non-metric connection with respect to the Galilean/Carrollian structures. The non-metricities are given by equation~\eqref{eq_Q_limit_Gal} and~\eqref{eq_Q_limit_Car} below. Since the Levi-Civita connection is torsion-free and metric, it seems natural to require that the reduced non-metricies of the Galilean/Carrollian connection obtained in the limit should be zero. In other words, we define the non-metricities such that the limit of the Levi-Civita connection of a Lorentzian metric is metric in the reduced variables. With such a requirement, we obtained a unique definition for the reduced non-metricities ($\hat \CQ_{\mu\nu}$, $\hat\CQ_\alpha{}^{\mu\nu}$, and $\check \CQ_{\mu}{}^{\nu}$, $\check\CQ_{\alpha\mu\nu}$) in both limits. Note that, in the Galilean case, this is only possible if $\T\tau\wedge\T\dd\T\tau =0$. We discuss this hypothesis in Section~\ref{sec_GH}.

In summary, we obtain/propose the following definitions for the reduced variables:

\paragraph{Galilean case:} imposing $\T\tau\wedge\T\dd\T\tau = 0$
\begin{itemize}
    \item For $\hat Q_{\mu\nu} = 0$ and $\hat Q_\alpha{}^{\mu\nu} = 0$, we define the {\it Galilean reduced torsion} as
    \begin{equation}
        \hat \CT^\alpha{}_{\mu\nu} \coloneqq T^\alpha{}_{\mu\nu} - 2v^\alpha  \omega_{\mu\nu}\,, \label{eq_def_reduced_Gal_torsion}
    \end{equation}
    and the identity~\eqref{eq_identities_Gal} becomes $\tau_\alpha \hat \CT^\alpha{}_{\mu\nu} = 0$. This quantity was also defined by \citet[]{2018_Bekaert_et_al} as $2\accentset{N}{U}^\alpha{}_{\mu\nu}$. Note that, as shown in Appendix~\ref{app_reduced_quantities_Gal}, assuming full metricity is not sufficient to get a unique definition for the reduced Galilean torsion (contrary to the reduced Carrollian torsion, as shown below). Therefore, the definition~\eqref{eq_def_reduced_Gal_torsion} remains a choice which we take to fit with \citep[]{2018_Bekaert_et_al}. The affine connection then takes the form
    \begin{align}
        \Gamma^\alpha_{\mu\nu} - {^{v,\tau}}\Gamma^\alpha_{\mu\nu} = \hat K^\alpha{}_{\mu\nu} + v^\alpha \omega_{\mu\nu} + 2\tau_{(\mu} \kappa_{\nu)\beta}h^{\alpha\beta}\,,
    \end{align}
    where we define the Galilean contorsion $\hat K^\alpha{}_{\mu\nu}\coloneqq -\hat \CT_{(\mu\nu)}{}^\alpha + \frac{1}{2} \hat \CT^\alpha{}_{\mu\nu}$ with respect to the reduced torsion.  
    \item For $T^\alpha{}_{\mu\nu} = 0$, we define the {\it Galilean reduced non-metricities} as
\begin{align}
    \hat \CQ_{\mu\nu} &\coloneqq \hat Q_{\mu\nu} + 2 \tau_\mu \omega_{\nu\sigma}v^\sigma,
\\
    \hat \CQ_\alpha{}^{\mu\nu} &\coloneqq \hat Q_\alpha{}^{\mu\nu} - 4 v^{(\mu}\omega^{\nu)}{}_\sigma v^{\sigma}\tau_\alpha\,,
\end{align}
and the identity~\eqref{eq_identities_Gal} becomes $\hat \CQ_{[\mu\nu]} = 0$ and $\tau_\nu \hat \CQ_\alpha{}^{\mu\nu} + \hat \CQ_\alpha{}^\mu = 0$. The affine connection then takes the form
    \begin{align}
        \Gamma^\alpha_{\mu\nu} - {^{v,\tau}}\Gamma^\alpha_{\mu\nu} = \hat L^\alpha{}_{\mu\nu} + 2v^\alpha v^\lambda\omega_{\lambda(\mu}\tau_{\nu)} + 2\tau_{(\mu} \kappa_{\nu)\beta}h^{\alpha\beta}\,,
    \end{align}
    where we define the Galilean distorsion $\hat L^\alpha{}_{\mu\nu} \coloneqq h^\alpha{}_\beta \hat \CQ_{(\mu\nu)}{}^\beta - \frac{1}{2} \hat \CQ^\alpha{}_{\mu\nu} - v^\alpha \hat \CQ_{(\mu\nu)}$ with respect to the reduced torsion.
\end{itemize}

Note that the present definitions of Galilean distorsion and contorsion do not include the Coriolis field, contrary to what was introduced in \citet{2024_WRV}. We feel it is a more natural way of defining these quantities as the presence of a Coriolis field is, in general, independent on the non-metricities and torsion.

\paragraph{Carrollian case:} 
\begin{itemize}
    \item For $\check Q_{\mu}{}^{\nu} = 0$ and $\check Q_{\alpha\mu\nu} = 0$, we define the the {\it Carrollian reduced torsion} as
    \begin{equation}
        \check \CT^\alpha{}_{\mu\nu} \coloneqq T^\alpha{}_{\mu\nu} - 2\tau_{[\mu}\Theta_{\nu]}{}^\sigma\,,
    \end{equation}
    and the identity~\eqref{eq_identities_Gal} becomes $v^\alpha\check \CT_{(\mu\nu)\alpha} = 0$. This quantity was also defined by \citet[Eq.\ 2.36]{2015_Hartong} as $2X^\alpha{}_{\mu\nu}$, and by \citet[Appendix 4]{2018_Bekaert_et_al} as $2\accentset{A}{U}^\alpha{}_{\mu\nu}$. The affine connection then takes the form
    \begin{align}
        \Gamma^\alpha_{\mu\nu} - {^{v,\tau}}\Gamma^\alpha_{\mu\nu} = \check K^\alpha{}_{\mu\nu} + \tau_\nu \Theta_{\mu\sigma}h^{\sigma\alpha} - v^\alpha \chi_{\mu\nu}\,,
    \end{align}
    where we defined the Carrollian contorsion $\check K^\alpha{}_{\mu\nu} \coloneqq -\check\CT_{(\mu\nu)}{}^\alpha + \frac{1}{2} \check\CT^\alpha{}_{\mu\nu}$.

    \item For $T^\alpha{}_{\mu\nu} = 0$, we define the the {\it Carrollian reduced non-metricities} as
\begin{align}
    \check \CQ_{\mu\nu} &\coloneqq \check Q_{\mu\nu} - \omega_{\mu\nu} - 2v^\sigma \omega_{\sigma(\mu} \tau_{\nu)}\,,
\\
    \check \CQ_\alpha{}^{\mu\nu} &\coloneqq \check Q_\alpha{}^{\mu\nu} - 4v^\lambda v^{(\mu} h^{\nu)\sigma} \omega_{\sigma(\alpha}\tau_{\lambda)}\,,
\end{align}
    and the identity~\eqref{eq_identities_Gal} becomes $v^\nu \check \CQ_{\alpha\mu\nu} = - \check \CQ_{\alpha\mu}$ and $v^\alpha \check \CQ_{\alpha\mu\nu}  = - 2 \check \CQ_{(\mu\nu)}$. The affine connection then takes the form
    \begin{align}
        \Gamma^\alpha_{\mu\nu} - {^{v,\tau}}\Gamma^\alpha_{\mu\nu} = \check L^\alpha{}_{\mu\nu} - v^\alpha \chi_{\mu\nu}\,,
    \end{align}
    where we defined the Carrollian distorsion $\check L^\alpha{}_{\mu\nu} \coloneqq -\check\CQ_{(\mu\nu)}{}^\alpha + \frac{1}{2} \check\CQ^\alpha{}_{\mu\nu}$.
\end{itemize}

There are two important points about these notions of Galilean and Carrollian distorsion and contorsion: (i) they depend on our choice of reduced quantities, and (ii) they can only be defined if the other one is zero. As pointed out in \cite{2025_Schwartz} for Galilean structures, in the general case with both metricities and torsion (and whether $\omega_{\mu\nu}=0$ holds or not) there is no meaningful way of defining distorsion and contorsion due to the identities~\eqref{eq_identities_Gal}. The same applies for the Carrollian case due to the identities~\eqref{eq_identities_Car}.\\

In summary, the reduced quantities, the contorsions and the distorsions are defined only if the affine connection is either metric or torsion-free. The advantage of using the reduced quantities is to be able to qualify a connection as being (with respect to reduced variables) both metric with respect to a Galilean or Carrollian structure and (with respect to reduced variables) torsion-free, while still having $\omega_{\mu\nu} \not= 0$ and $\Theta_{\mu\nu} \neq 0$. As will be discussed later, the reduced torsion and reduced non-metricities are the quantities that should be considered in the discussion about the a trinity formulation of Carrollian (or Galilean) gravity.

One important disadvantage of these quantities is that they are not invariant under Galilean or Carrollian boosts, i.e.\ they depend on the choice of  $v^\mu$ in the former case, and on the choice of $\tau_\mu$ in the latter case. Consequently, in general, an affine connection is (with respect to reduced variables) metric or torsionless only with respect to a specific observer or Ehresmann connection.

\section{Galilean and Carrollian limits}
\label{sec_Limits}

In this section we discuss the general decomposition derived in the previous section in the context of the Galilean and Carrollian limits of the Levi-Civita connection of a Lorentzian metric. The results of this section are summarized in Table~\ref{tab_limits}.

\subsection{Galilean limit}

The Galilean limit of a Lorentzian metric $\T g$ is defined with the following leading order~(LO) ansatz (for $\lambda \coloneqq 1/c$) \citep{1976_Kunzle}:
\begin{align}\label{eq_NR_limit}
	\tayll{g}^{\mu\nu}	= h^{\mu\nu} + \lambda^2 \, \tayl{2}{g}^{\, \mu\nu} + \bigO{\lambda^4}, \quad \tayll{g}_{\mu\nu}	= -\tfrac{1}{\lambda^2}\tau_\mu \tau_\nu + \tayl{0}{g}_{\mu\nu} + \bigO{\lambda^2}. 
\end{align}
From the identity relation $g^{\mu\alpha} g_{\alpha\mu} = \delta^\mu_\nu$, the next-to-leading orders~(NLO) take the form
\begin{align}
	\tayl{2}{g}^{\,\mu\nu}	= -v^\mu v^\nu + k^{\mu\nu}\,,
    \quad \tayl{0}{g}_{\mu\nu}	= \gamma_{\mu\nu} -2\phi\tau_\mu\tau_\nu\,,
\end{align}
with $\tau_\mu k^{\mu\nu} = 0$, and where $\tau_\mu v^\mu = 1$ and $\gamma_{\mu\nu}$ is the projector orthogonal to $v^\mu$.

From the ansatz~\eqref{eq_NR_limit}, the LO and NLO of the Levi-Civita connection ${{^g}\Gamma}^\alpha_{\mu\nu}$ of the Lorentzian metric $\tayll{\T g}$ are
\begin{align}
	\tayl{-2}{{{^g}\Gamma}}^\alpha_{\mu\nu}
		&= -2\tau_{(\mu} \omega_{\nu)}{}^\alpha\,, \\
    \tayl{0}{{{^g}\Gamma}}^\alpha_{\mu\nu}
		&= {^{v,\tau}{\Gamma}}^\alpha_{\mu\nu} + 2\left(-2\phi h^{\alpha\sigma} + v^\alpha v^\sigma - k^{\alpha\sigma}\right)\tau_{(\mu}\omega_{\nu)\sigma} + \tau_\mu \tau_\nu h^{\alpha\sigma}\partial_\sigma\phi\,.
\end{align}
With respect to the (torsion-free) connection at zeroth order, we get
\begin{align}
    \hat Q_{\mu\nu}
        &= \omega_{\mu\nu} - 2\tau_{(\mu}\omega_{\nu)\sigma} v^\sigma\,, 
            \quad \hat Q_\alpha{}^{\mu\nu} = 2 v^{(\mu} \omega^{\nu)}{}_\alpha + 2\tau_\alpha\left[v^{(\mu}\omega^{\nu)}{}_\sigma v^\sigma - k^{\sigma(\mu} \omega^{\nu)}{}_\sigma\right],\label{eq_Q_limit_Gal}\\
    \kappa_{\mu\nu}
        &= \tau_{[\mu} \partial_{\nu]}\phi - 2\phi \omega_{\mu\nu} + k^\sigma{}_{[\mu} \omega_{\nu]\sigma} + k^\sigma{}_{[\mu} \tau_{\nu]} v^\lambda \omega_{\lambda\sigma}\,. \label{eq_Coriolis_field_limit}
\end{align}
The non-metricity $\hat Q_\alpha{}^{\mu\nu}$ depends not only on $\omega_{\mu\nu}$ but also on $k^{\mu\nu}$ which is of NLO. However, if $\T\tau\wedge\T\dd\T\tau=0$, the pure spatial part of $\omega_{\mu\nu}$ is zero, the term $k^{\sigma(\mu} \omega^{\nu)}{}_\sigma$ vanishes, and the non-metricity only depends on $\omega_{\mu\nu}$. This makes it possible to define a reduced non-metricity fulfilling the conditions given in Section~\ref{sec_Reduced}, i.e. depending solely on $\omega_{\mu\nu}$. The Frobenius condition on $\T\tau$ is generally assumed or obtained from the limit of general relativity if matter only contributes from the ($-4$) order in the limit (see, e.g., \cite{2020_Hansen_et_al}). In Section~\ref{sec_GH}, we show that considering global hyperbolicity of the Lorentzian {manifold}, before any field equation, already leads to the Frobenius condition on $\T\tau$.\\

\begin{remark}
In the literature on the Galilean limit, a torsionful metric connection is usually considered rather than a non-metric torsion-free one. For instance, \citet[][]{2020_Hansen_et_al} use ${{^{v,\tau}\Gamma}^\alpha_{\mu\nu} + v^\alpha\omega_{\mu\nu}}$ to express the equations obtained in the limit of general relativity. The reduced torsion of this connection is zero. Therefore, in this approach, we can qualify the connection as being (with respect to reduced variables) metric and torsion-free, while $\tayl{0}{\Gamma}^\alpha_{\mu\nu}$ is (with respect to reduced variables) metric and torsion-free. 
Both descriptions lead to field equations obtained from the limit of the Einstein-Hilbert Lagrangian which are considerably more complicated (see Eq.~(3.39) to (3.41) in \citep[][]{2020_Hansen_et_al}), and less explicit in the physics they encode, than the well-known Newton-Cartan equations (valid for $\omega_{\mu\nu} = 0$). In our opinion, this suggests that a better way of describing the field equations when $\omega_{\mu\nu} \not=0$ would be to come back to a classical spatial description with a 3+1 decomposition. This approach has the advantage of working with a uniquely defined (spatial) symmetric and metric connection. 
The same remark will apply to the Carrollian limit when~$\Theta_{\mu\nu} \not=0$.
\end{remark}

\subsection{Carrollian limit}

The Carrollian limit of a Lorentzian metric $\T g$ is defined with the following LO ansatz (for $\epsilon \coloneqq c$) \citep{1998_Dautcourt}:
\begin{align}\label{eq_UR_limit}
	\tayle{g}^{\mu\nu}	= -\frac{1}{\epsilon^2} v^\mu v^\nu + \tayl{0}{g}^{\, \mu\nu} + \bigO{\epsilon^2}, \quad \tayle{g}_{\mu\nu}	= \gamma_{\mu\nu} + \epsilon^2\, \tayl{2}{g}_{\mu\nu} + \bigO{\epsilon^4}\,. 
\end{align}
From the identity relation $g^{\mu\alpha} g_{\alpha\mu} = \delta^\mu_\nu$, the NLO take the form
\begin{align}
	\tayl{0}{g}^{\,\mu\nu}	= h^{\mu\nu} - 2\phi v^\mu v^\nu\,,
    \quad \tayl{2}{g}_{\mu\nu}	= - \tau_\mu \tau_\nu + k_{\mu\nu}\,,
\end{align}
with $v^\mu k_{\mu\nu} = 0$, and where $\tau_\mu v^\mu \coloneqq 1$ and $h^{\mu\nu}$ is the projector orthogonal to $\tau_\mu$.

From the ansatz~\eqref{eq_UR_limit}, the LO and NLO of the Levi-Civita connection ${{^g}\Gamma}^\alpha_{\mu\nu}$ of the Lorentzian metric $\tayle{\T g}$ are
\begin{align}
	\tayl{-2}{{{^g}\Gamma}}^\alpha_{\mu\nu}
		&= -v^\alpha \, \Theta_{\mu\nu}\,, \label{eq_conn_2} \\
    \tayl{0}{{{^g}\Gamma}}^\alpha_{\mu\nu}
		&= {^{v,\tau}{\Gamma}}^\alpha_{\mu\nu} + v^\alpha \left[\Lie{\T v} k_{\mu\nu} + 2\phi \, \Theta_{\mu\nu} + v^\sigma \left(\tau_{\mu} \partial_{[\nu} \tau_{\sigma ]} + \tau_{\nu} \partial_{[\mu} \tau_{\sigma ]}\right)\right].
\end{align}

With respect to the (torsion-free) connection at zeroth order, we get
\begin{align}
    \check Q_{\mu}{}^{\nu}
        &= \Theta_\mu{}^\nu \,, \quad \check Q_{\alpha\mu\nu}
        = -2\tau_{(\mu} \Theta_{\nu)\alpha}\,, \label{eq_Q_limit_Car} \\
    \chi_{\mu\nu}
        &= -\frac{1}{2}\Lie{\T v}k_{\mu\nu} - 2\phi \Theta_{\mu\nu} - v^\sigma \left(\tau_{\mu} \partial_{[\nu} \tau_{\sigma ]} + \tau_{\nu} \partial_{[\mu} \tau_{\sigma ]}\right). \label{eq:chi}
\end{align}
Therefore, we see that the Carrollian limit of a symmetric (Levi-Civita) Lorentzian connection is, at zeroth order, a torsion-free connection with zero reduced non-metricities.

\begin{table}[t]
\centering\small
\caption{\small Properties of a Lorentzian connection in the limit. In each case, the NLO orders of the metrics are written with respect to the (unique) $v^\mu$ and $\tau_\mu$ for which no shift is present.\label{tab_limits}}
\renewcommand{\arraystretch}{1.5}
\newcommand{\vspacetable}[1]{\multicolumn{2}{c}{} \vspace{#1}\\}
\begin{tabular}{>{\centering\arraybackslash}m{\columnwidth*9/50-2\tabcolsep}|p{\columnwidth*19/50-2\tabcolsep}|p{\columnwidth*22/50-2\tabcolsep}}
	\hline\hline 
    
\bf\centering  Quantities
    &\centering \bf Galilean structure
    & \centering \bf Carrollian structure \arraybackslash\\
\hline\hline 

LO Ansatz
    & $\begin{aligned}
	   \tayl{0}{g}^{\, \mu\nu} = h^{\mu\nu} \ \ ;\ \ \tayl{-2}{g}_{\mu\nu} = -\tau_\mu\tau_\nu
    \end{aligned}$
    & \addstackgap{$\begin{aligned}
	   \tayl{-2}{g}^{\ \, \mu\nu}	= -v^\mu v^\nu \ \ ;\ \ \tayl{0}{g}_{\mu\nu} = \gamma_{\mu\nu}
    \end{aligned}$}
    \\
\hline

NLO
    & $\begin{aligned}
	  \tayl{2}{g}^{\,\mu\nu} &= -v^\mu v^\nu + k^{\mu\nu} \\
	  \tayl{0}{g}_{\mu\nu} &= \gamma_{\mu\nu} - 2\phi \tau_\mu\tau_\nu 
    \end{aligned}$
    & \addstackgap{$\begin{aligned}
	   \tayl{0}{g}^{\,\mu\nu} &= h^{\mu\nu} - 2\phi v^\mu v^\nu \\
	  \tayl{2}{g}_{\mu\nu} &= -\tau_\mu \tau_\nu + k_{\mu\nu} 
    \end{aligned}$}
    \\
\hline

\centering Decomposition of 
    
    $\tayl{0}{{{^g}\Gamma}}^\alpha_{\mu\nu} - {^{v,\tau}}\Gamma^\alpha_{\mu\nu}$
    & $\begin{aligned} 
        &T^\alpha{}_{\mu\nu} = 0, \ \hat\CQ_{\mu\nu} = 0\ ; \ \hat\CQ_\alpha{}^{\mu\nu} = 0 \\
        &\kappa_{\mu\nu} = \tau_{[\mu} \partial_{\nu]}\phi - 2\phi \omega_{\mu\nu}
    \end{aligned}$
 
    & \addstackgap{$\begin{aligned} 
        T^\alpha{}_{\mu\nu} &= 0, \ \check\CQ_{\mu}{}^\nu = 0\ ; \ \check\CQ_{\alpha\mu\nu} = 0 \\
        \chi_{\mu\nu} &= -\frac{1}{2}\Lie{v}k_{\mu\nu} - 2\phi \Theta_{\mu\nu} - 2v^\sigma \tau_{(\mu} \omega_{\nu)\sigma}
    \end{aligned}$} \\
\hline

\centering \vspace{.cm} Constraints from global hyperbolicity
    &   \begin{tabular}{@{}l@{}}
            (i) $\T\tau\wedge\T\dd\T\tau = 0$ \\
                $\quad\ $ and $\T \tau = N \T \dd t$ holds globally\\ 
            (ii) $\exists$ complete $\tilde v^\mu$ with $\tau_\mu \tilde v^\mu = 1$ \\
            \\
            (iii) $h^{\mu\alpha}\kappa_{\alpha\beta}h^{\beta\nu} = 0$, \\
                $\quad\ $ i.e. no global 3-rotations
        \end{tabular}
     &  \vspace{-2.25cm}\begin{tabular}{@{}l@{}}
            (i) $\T v$ is a complete vector field \\
            \\
            (ii) $\exists \, \tilde\tau_\mu$ with $\tilde\tau_\mu v^\mu = 1$ \\
                $\quad\ $ such that $\tilde{\T\tau} = N \T \dd t$ holds globally \\
            (iii) no dual constraint on $\chi_{\mu\nu}$
        \end{tabular}\\
\hline
\end{tabular}
\end{table}

\subsection{The role of global hyperbolicity}\label{sec_GH}

\subsubsection{Motivations}

Additional constraints (e.g. $h^{\mu\sigma}\omega_{\sigma\kappa}h^{\kappa\nu} = 0$ or $\Theta_{\mu\nu} = 0$) on the Galilean/Carrollian structures arising from the limit of a Lorentzian structure can be obtained by considering the Einstein equation with a specific choice of energy-momentum tensor.

In particular, in the Galilean case, it is generally assumed that the matter Lagrangian has no $(-6)$ order, which implies $h^{\mu\sigma}\omega_{\sigma\kappa}h^{\kappa\nu} = 0$ and $\T\tau$ is foliation forming, hence inducing an absolute standard of simultaneity (see, e.g.~\cite{1997_Dautcourt, 2020_Hansen_et_al}). Therefore, locally we have {${\T\tau = N\T\dd t}$}, where $N$ is a non-vanishing scalar field, and $t$ a scalar field. But because $t$ is only defined locally, in general the leaves of the foliation cannot be labelled by a scalar field defined on $\mR$. In other words, while $h^{\mu\sigma}\omega_{\sigma\kappa}h^{\kappa\nu} = 0$ implies absolute simultaneity, the Newtonian time is not guaranteed to run indefinitely. For example, this situation arises if the spacetime is $\CM = \mR\times\Sigma$ with $\Sigma$ a closed 3-manifold and if the leaves of the $\T\tau$-foliation are not isomorphic to $\Sigma$.

Since the assumption of infinite Newtonian time, common in the literature on Galilean structures and the Galilean limit, is not guaranteed by the condition $h^{\mu\sigma}\omega_{\sigma\kappa}h^{\kappa\nu} = 0$, then the property that $\T\tau = N\T\dd t$ holds globally must be treated as an additional hypothesis, not implied by either the limit or the extremalisation of the action.

The goal of this section is to show that a natural geometric condition on the Lorentzian structure from which the Galilean structure arises in the limit, namely global hyperbolicity, implies that $\T\tau = N\T\dd t$ holds globally prior to any considerations on the Lagrangian of matter. Hence this guarantees that the Newtonian time is defined on $\mR$. We will also derive other properties that are implied by global hyperbolicity, both in the Galilean and the Carrollian case. The results are summarised in the final line of Table~\ref{tab_limits}.

\subsubsection{Derivation}

Let us consider the limit of timelike vector fields $\tayll{u}^\mu$ (Galilean case) and $\tayle{u}^\mu$ (Carrollian case). Using $\tayll{u}^\nu \tayll{u}^\mu \tayll{g}_{\mu\nu} = -c^2 = -\lambda^{-2}$ and $\tayle{u}^\nu \tayle{u}^\mu \tayle{g}_{\mu\nu} = -\epsilon^2$, the leading order of both $\tayll{u}^\mu$ and $\tayll{u}_\mu$ (and both $\tayle{u}^\mu$ and $\tayle{u}_\mu$) are constrained.

In the Galilean limit, we have
\begin{align}
    \tayll{u}^\mu = \tayl{0}{u}^{\,\mu} + \bigO{\lambda^{2}}, \quad \tayll{u}_\mu = -\frac{1}{\lambda^2} \tau_\mu + \bigO{\lambda^0}, \label{eq_LO_u_Gal}
\end{align}
with $\tayl{0}{u}^{\,\mu} \tau_\mu = 1$. The sign convention is so that $\tayl{0}{u}^{\, \mu}$ is future directed.

In the Carrollian limit, we have
\begin{align}
    \tayle{u}^\mu = v^\mu + \bigO{\epsilon^2}, \quad \tayle{u}_\mu = -\epsilon^2 \,\tayl{2}{u}_{\mu} + \bigO{\epsilon^{4}}, \label{eq_LO_u_Car}
\end{align}
with $\tayl{2}{u}_\mu v^\mu = 1$.

Global hyperbolicity of $(\CM, \T g)$ ensures, in particular, that there exists a foliation forming timelike 1-form $\T n$ that can be written as $\T n = N\T{\dd} t$ globally. Additionally we can choose $\T n$ such that the dual vector $\T n^\sharp := \T g(\T n, \cdot)$ is complete and nowhere vanishing.\footnote{To ensure completeness of the vector field $\T n^\sharp$ it is sufficient to choose the 1-form  $\T n$ such that $\int_\mR N\dd t = \infty$, which can always be done without loss of generality on the metric $\T g$.}

We consider the Taylor series of such a timelike 1-form/vector. Because we assume analyticity, then each order of $\tayll{n}^\mu$ and $\tayle{n}^\mu$ has to be a complete vector field, and each order of $\tayll{n}_\mu$ and $\tayle{n}_\mu$ has the form $N\T \dd t$ globally. This holds in particular for the leading orders given in equations~\eqref{eq_LO_u_Gal} and~\eqref{eq_LO_u_Car}. Therefore, global hyperbolicity imposes:
\begin{itemize}
    \item In the Galilean limit:  that $\T\tau = N\T\dd t$ holds globally, and there exists a complete vector field $u^\mu$ such that $u^\mu \tau_\mu = 1$.
    \item In the Carrollian limit: that $v^\mu$ is a complete vector field, and there exists a 1-form $\T\tau = N\T\dd t$ such that $\tau_\mu v^\mu = 1$. This implies in particular that there exists a foliation never tangent to $v^\mu$.
\end{itemize}

Therefore, while in the Carrollian case, there are no direct constraints from global hyperbolicity on the connection $\tayl{0}{\Gamma}^\mu_{\alpha\beta}$ in the limit, in the Galilean case, because $\T\tau$ is foliation forming, then the Coriolis field has no spatial part as can be seen by spatially projecting relation~\eqref{eq_Coriolis_field_limit}. To be more precise, the Coriolis field is not unique as it depends on the choice of observer with $v^\mu$. Therefore, for a general observer, the Coriolis field associated with that observer will have a spatial part. Global hyperbolicity only ensures that there exist observers with respect to which the spatial part of their Coriolis field is zero. Physically, this ensures the existence of irrotational observers.

\section{Geometric trinities of gravitation}\label{sec:trinity}

One topic which has been the focus of significant and sustained focus in the recent literature is the so-called `geometric trinity' of relativistic gravitational theories, in which (a) curvature degrees of freedom in GR are traded either for torsion degrees of freedom (as for the `teleparallel equivalent of GR' (TEGR)) or non-metricity degrees of freedom (as for the `symmetric teleparallel equivalent of GR' (STEGR)), and (b) the resulting theories are shown to have actions which are equivalent up to a boundary term, and as such are dynamically equivalent. (See \cite{2019_Jimenez_et_al} for a review of the geometric trinity.) What we wish to consider in this section, in a unified way, are the Galilean and Carrollian limits of the geometric trinity.

The Galilean limit of a STEGR and a TEGR connection with $\omega_{\mu\nu} = 0$ was derived in \citet{2024_WRV} and \citet{2023_Schwartz}, respectively. In these approaches, the leading order of the connection is assumed to be the zeroth order. While this assumption is not possible for the Levi-Civita connection without restriction on $\omega_{\mu\nu}$ or $\Theta_{\mu\nu}$, it can be considered if the connection is not Levi-Civita without loss of generality. In what follows, we keep the assumptions ${^{\rm STEGR}\Gamma^\alpha_{\mu\nu}} = \tayl{0}{\Gamma}^\alpha_{\mu\nu} + \bigO{\epsilon^2\text{ or }\lambda^2}$ and ${^{\rm TEGR}\Gamma^\alpha_{\mu\nu}} = \tayl{0}{\Gamma}^\alpha_{\mu\nu} + \bigO{\epsilon^2\text{ or }\lambda^2}$. In both cases, the flatness of the full connection implies the flatness of the zeroth order connection.

The Galilean limit of STEGR was studied in \cite{2024_WRV} for $\omega_{\mu\nu} =0$. As shown in that paper, the connection at zeroth order is torsion-free and non-metric with respect to the Galilean structure obtained in the limit from the ansatz~\eqref{eq_NR_limit}. That result remains the same if $\omega_{\mu\nu} \not=0$, with the additional property that the reduced non-metricities of $\tayl{0}{\T\nabla}$ are non-zero. In other words, the non-metricities are not due only to $\omega_{\mu\nu}$ but encode additional physics. Therefore,  the Galilean limit of STEGR is characterized by a flat, torsion-free and (with respect to reduced variables) non-metric connection, i.e. $R^\mu{}_{\nu\alpha\beta} = 0$, $T^\mu{}_{\alpha\beta} = 0$, $\hat \CQ_{\mu\nu} \not=0$ and $\hat\CQ_\alpha{}^{\mu\nu} \not=0$ for~$\tayl{0}{\T\nabla}$.

\begin{figure}[t]
    \centering
     \includegraphics{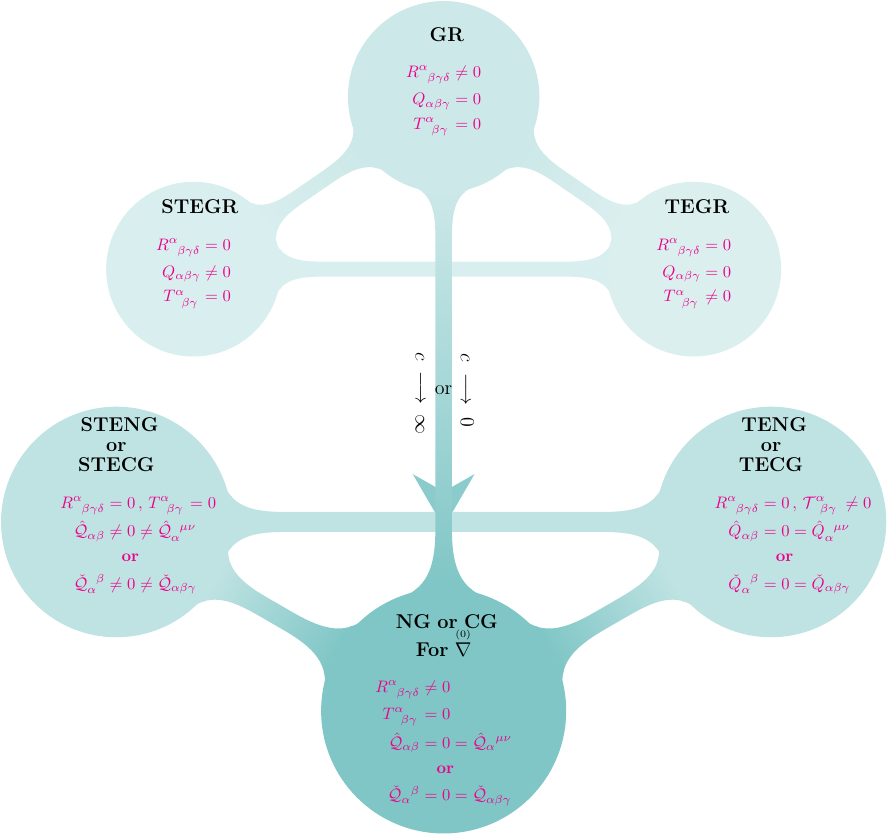}
    \caption{The relativistic geometric trinity of gravitational theories (top), and its Galilean/Carrollian limits (bottom).}
    \label{fig_trinity}
\end{figure}

In the Carrollian limit, the same applies. The zeroth-order connection is torsion free, and from $\tayle{\nabla}_\alpha \tayle{g}_{\mu\nu}\not=0$, we get $\check Q_\mu{}^\nu = \tayl{0}{\nabla}_\mu v^\nu \not=0$ and $\check Q_{\alpha\mu\nu} = \tayl{0}{\nabla}_\alpha \gamma_{\mu\nu} \not=0$. The reduced non-metricities are in general also non-zero which contrasts with the limit of the Levi-Civita connection for which the non-metricities were only due to $\Theta_{\mu\nu}\not=0$. Therefore,  as in the the Galilean limit, the Carrollian limit of STEGR is characterized by a flat, torsion-free and (with respect to reduced variables) non-metric connection, i.e. $R^\mu{}_{\nu\alpha\beta} = 0$, $T^\mu{}_{\alpha\beta} = 0$, $\check \CQ_{\mu}{}^{\nu} \not=0$ and $\check\CQ_{\alpha\mu\nu} \not=0$~for $\tayl{0}{\T\nabla}$.

The Galilean limit of TEGR was derived by \citet{2023_Schwartz}, assuming $\omega_{\mu\nu} = 0$. In the limit, the connection at zeroth order is metric, flat and torsionful. Including $\omega_{\mu\nu} \not= 0$ does not change this result. As for STEGR, the physics not already encoded in $\Theta_{\mu\nu}$ is encoded in the reduced quantity, here the reduced torsion. The same applies for the Carrollian limit for which the zeroth order of the connection is flat, metric and (with respect to reduced variables) torsionful.

In summary:
\begin{itemize}
    \item[(i)]The Galilean/Carrollian limit of STEGR for general structures with $\omega_{\mu\nu} \not=0$ and $\Theta_{\mu\nu} \not=0$ is characterized by a flat, torsion-free connection with reduced non-metricities.
    \item[(i)]The Galilean/Carrollian limit of TEGR for general structures with $\omega_{\mu\nu} \not=0$ and $\Theta_{\mu\nu} \not=0$ is characterized by a flat, metric connection with reduced torsion.
\end{itemize}
This is schematized in Figure~\ref{fig_trinity}.

A remark on variational principles. Although the relativistic geometric trinity is often presented at the level of actions (which differ by boundary terms---see \cite{2017_Jimenez_et_al}), the non-relativistic geometric trinity of gravitational theories was presented in \cite{2024_WRV} entirely at the level of the equations of motion, because there are by now well-known obstructions to formulating Newton--Cartan theory (and for that reason also its teleparallel and symmetric teleparallel equivalents) in terms of an action principle (although it is possible to construct a more general version of Newton--Cartan theory, known as `Type II' Newton--Cartan theory, which does admit an action principle---see \cite{2019_Hansen_et_al_b}). In principle, one could consider \emph{general} Galilean and Carrollian Lagrangians, functions of the reduced variables and (respectively) the tensors ($\omega_{\mu\nu}, \kappa_{\mu\nu}$) and ($\Theta_{\mu\nu}, \chi_{\mu\nu}$); however, at such a level of generality such a discussion would not obviously be enlightening, and for this reason we eschew an explicit presentation here. Such Lagrangians would be the appropriate starting point for considering the dynamics of Galilean and Carrollian metric-affine theories of gravitation (following the lead of \cite{1995_Hehl_et_al} in the relativistic case), which would be substantially more general than either the Galilean or Carrollian geometric trinities; exploring such dynamics would be to take up Schwartz' injunction in \cite{2025_Schwartz} to investigate Galilean (and in our case also Carrollian) metric-affine theories of gravitation.

Completionism aside, there are various reasons why it is worth considering something like a Carrollian geometric trinity. One obvious motivation is the following: \cite{2017_Oshita_et_al} have shown that it is easier to model boundary phenomena (associated with e.g.\ the physics of black hole boundaries) when working with (S)TEGR rather than GR---for the former has a well-posed variational problem, whereas the latter does not (for further discussion of this, see \cite{2023_Wolf_et_al_a}). In addition, Carrollian structures find their most notable physical application in the context of asymptotics and boundaries---for it is by now well-known that the Carroll group is closely related to the BMS group (see \cite{2014_Duval_et_al}). Bringing these points together, it is not unreasonable to expect that something like (S)TECG might find application in the context of the study of such boundary phenomena.

\section{Compatible Galilean and Carrollian connections}\label{sec:com-connections}

On the basis of the constructions which we have presented up to this point in this article, we consider now a follow-on---but nonetheless interesting---issue. Rather than considering the relation between an affine connection and a single structure—either Galilean, Carrollian—we can wonder what kind of constraints we obtain if we consider a torsion-free connection which is compatible with \emph{both} Galilean and Carrollian structures. In other words, what constraints on $\T\nabla$, $(\tau_\mu, h^{\mu\nu})$, $(v^\mu, \gamma_{\mu\nu})$ do we get if
\begin{align}
    \hat Q_{\mu\nu}=0, \quad \hat Q_\mu{}^{\alpha\beta}=0, \quad \check Q_\mu{}^{\nu} = 0, \quad \check Q_{\mu\alpha\beta} = 0, \quad \text{and} \quad T^\mu{}_{\alpha\beta} = 0 \,? \label{eq_conditions}
\end{align}

These conditions imply that $\T v$ is a collineation vector field for $\T h$, $\T\tau$, $\T\gamma$, the connection $\T\nabla$ and $\T{\rm Riem}$, i.e.
\begin{align}
    \begin{split}
    &\Lie{\T v} \T \tau = 0, \qquad \Lie{\T v} \T h = 0, \qquad \Lie{\T v} \T \gamma = 0, \\
    &\Lie{\T v} \Gamma^\alpha_{\mu\nu} \coloneqq  \nabla_\mu \nabla_\nu v^\alpha - R^\alpha{}_{\sigma\mu\nu}v^\sigma = 0, \qquad \Lie{\T v} R^\alpha{}_{\beta\mu\nu} = 0.
    \end{split}\label{eq_Lie}
\end{align}
Additionally, $\T\dd\T\tau = 0$ and $\T\tau$ defines a foliation.

From $\nabla_\mu \left(\tau_\nu v^\nu\right) = 0$, we get $\tau_\mu v^\nu = \text{const}$. This means that the Carrollian and Galilean time metrics are either dual if $\tau_\mu v^\nu \not =0$ or orthogonal with $\tau_\mu v^\nu = 0$. The latter case is a peculiar situation which—if $\T\tau$ defines a closed spatial foliation as is considered in Newtonian cosmology—implies that the flows of $\T v$ would be tangent to that foliation and therefore periodic. Therefore, while not forbidden by~\eqref{eq_conditions}, the case $\tau_\mu v^\nu = 0$ involves a property of the time metric of the Carrollian structure usually discarded. In what follows, we consider the case $\tau_\mu v^\nu \not= 0$, and, without loss of generality, we assume $\tau_\mu v^\nu = 1$. 

This does not yet mean that $(\tau_\mu, h^{\mu\nu})$ and $(v^\mu, \gamma_{\mu\nu})$ are dual. Indeed, the relation ``$\gamma_{\mu\alpha} h^{\alpha\nu} = \delta^\nu_\mu - v^\nu \tau_\mu$'' does not necessarily hold. This means that, while $\gamma_{\mu\nu}$ induces a spatial metric $\gamma_{ij}$ on the leafs of the $\T\tau$-foliation, that metric is not necessarily equivalent to the metric $h_{ij}$ induced by the Galilean structure. 
However, the spatial connection $\T{\hat D}$ induced by $\T\nabla$ and $\T h$ on the $\T\tau$-foliation is compatible with $\T\gamma$ (i.e. $h^{\alpha\kappa} h^{\mu\sigma} h^{\nu\lambda} \nabla_\kappa \gamma_{\sigma\lambda} = 0$), and reversely the spatial connection $\T{\check D}$ induced by $\T\nabla$ and $\T \gamma$ on the $\T\tau$-foliation is compatible with $\T h$ (i.e. $\gamma_{\alpha\beta}h^{\beta\kappa} \gamma_{\mu\sigma} \gamma_{\nu\lambda} \nabla_\kappa h^{\sigma\lambda} = 0$). 
This means that while $\T h$ and $\T \gamma$ are not necessarily inducing the same spatial metric on the $\T\tau$-foliation, their spatial Levi-Civita connection are equivalent. There is therefore a unique spatial Ricci curvature that is induced by $\T\nabla$, $\T h$ or $\T\gamma$ on the $\T\tau$-foliation.

Continuing with the constraints that~\eqref{eq_conditions} impose, $\nabla_\mu v^\nu = 0$ implies the existence of an observer with respect to the Galilean structure which is irrotational (i.e. ${^{\T v}\Omega}^{\mu\nu} \coloneqq h^{\kappa[\mu} \nabla_\kappa v^{\nu]} = 0$), non-expanding (i.e. ${^{\T v}\Theta}^{\mu\nu} \coloneqq h^{\kappa(\mu} \nabla_\kappa v^{\nu)} = 0$) and non-accelerating (i.e. ${^{\T v}a}^{\mu} \coloneqq v^\nu \nabla_\nu v^{\mu} = 0$). These are all the dynamical degrees of freedom of the connection. Using a 3+1-projection along $\T\tau$, $\T v$, $\T h$ and $\T\gamma$ of the Riemann tensor of $\T\nabla$ (see, e.g., \cite{2021_Vigneron}), and choosing a basis $\{\T v, \T{e}_i \}$ adapted to the $\T\tau$-foliation (i.e.\ $\T n(\T e_i) = 0$), the Riemann tensor takes the form
\begin{align}
    R^\alpha{}_{\beta\mu\nu} = \delta^{\alpha}_{ i} \delta^{ j}_{ \beta} \delta^{ k}_{ \mu} \delta^{ l}_{ \nu}\, \CR^{ i}{}_{ j  k  l}\,,
\end{align}
where $\CR^{ i}{}_{ j  k  l}$ is the spatial curvature induced by $\T h$ (or $\T\gamma$) on the $\T\tau$-foliation, where Latin indices stand for the spatial directions. Using the last constraint of~\eqref{eq_Lie}, this implies that the spatial Riemann tensor is static.

In summary, the freedom left on a symmetric affine connection compatible with a Galilean and Carrollian structures (that are not orthogonal to each other) is a static spatial curvature. In particular, in a basis adapted to the $\T\tau$-foliation, the Ricci tensor has the form
\begin{align}
    R_{\mu\nu} =
        \begin{pmatrix}
            0 & 0 \\
            0 & \CR_{ij}
        \end{pmatrix}.\label{eq_R_ij}
\end{align}

That connection is also compatible with a (Lorentzian or Riemannian) metric of the form $g_{\mu\nu} = {\rm diag}\left(\pm 1, h_{ij}\right)$ in the adapted basis $\{\T v, \T{e}_i \}$. In that case, $\T v$ is a Killing vector for the metric. This shows that the above connection is actually compatible with three structures: Galilean, Carrollian and Riemannian/Lorentzian.

Interestingly, this kind of connection is exactly what was introduced as the ``reference connection'' $\T{\bar\nabla}$ in \citet{2024_Vigneron} to ensure the existence of the Galilean limit for any spatial topology. In this framework, $\T{\bar\nabla}$ is related to the connection of the universal covering space, a topological property of the spatial slices. Therefore, while highly specific, the type of connection obtained from the constraints~\eqref{eq_conditions} finds application in cosmology when non-Euclidean topologies are considered. Quite apart from potential applications to physics, however, it is certainly interesting to see that simultaneously insisting that a connection be compatible with Carrollian and Galilean structures---what one might consider to be quite stringent conditions---still leaves significant freedom for the resulting connection.

\section{Conclusion}\label{sec:close}

In this article, following the lead of Schwartz \cite{2025_Schwartz} for the Galilean case, we have presented for the first time the general form of a Carrollian connection with both torsion and non-metricity; we have also compared our results with the form of a general Galilean connection in a completely general way. Having done so, we have shown how these results can be militated in order to construct an ultra-relativistic geometric trinity of gravitational theories, and have considered the geometry and physics of connections which are simultaneously compatible with both Galilean and Carrollian structures.

Building on this work, many further interesting questions for future pursuit arise. Here are four such questions:
\begin{enumerate}
    \item In the Galilean case, it is known that suitable gauge fixing (essentially that the spatial torsion vanish---see \cite{2023_Schwartz, 2018_Read_et_al, 2024_WRV}) suffices to recover standard, potential-based Newtonian gravity. Hence, one might wonder about the conditions under which this would also be possible in the Carrollian case. \emph{Prima facie}, this is not likely to be as straightforward as in the Galilean case, where in fact recovery of a gravitational potential relies crucially on the asymmetry of the Coriolis form; a different strategy will have to be pursued, and the physical justification for that strategy remains to be seen.\footnote{In \cite[p.\ 35]{2018_Bekaert_et_al}, \eqref{eq:theta} is described as a `Carrollian potential'; this, however, is not obviously what one is after if one is thinking in analogy with the Newtonian gravitational potential.}

    \item In \cite{2023_March_et_al}, it is shown that the common structure of the non-relativistic geometry trinity of gravitational theories is a theory with only a standard of absolute rotation and no standard of non-rotational acceleration, known as `Maxwell gravitation'. Building upon our above construction of an ultra-relativistic geometric trinity, what would the common structure of this new trinity be, and (if it exists) does this common structure qualify as an interesting physical theory in its own right?

    \item Schwartz \cite[\S4]{2025_Schwartz} goes on to offer a fibre bundle perspective on general Galilean connections---could an analogous gauge-theoretic perspective on general Carroll connections likewise be provided?

    \item  We have already mentioned above that one could use the results of this article to investigate general Galilean and Carrollian metric-affine theories of gravitation. What we have provided in this article is a unified geometrical scaffolding within which to construct and appraise such theories (and, crucially, their solutions); investigating the details of such theories is surely an important next step.
\end{enumerate}

\section*{Acknowledgments}

\noindent QV is supported by the Polish National Science Centre under grant No. SONATINA 2022/44/C/ST9/00078. \\

\noindent HB was funded by the Austrian Science Fund (FWF) [Grant DOI: 10.55776/J4803].
For open access purposes, the authors have applied a CC BY public copyright license to any author-accepted manuscript version arising from this submission. \\

\noindent We thank Eleanor March and Will Wolf for valuable discussions.

\appendix

\section{General reduced quantities}\label{app_reduced_quantities}

In this appendix, we present the (non-unique) general approach to define reduced quantities without considering the Galilean or Carrollian limit.

\subsection{Galilean case}\label{app_reduced_quantities_Gal}

The goal is to define the reduced quantities $\hat\CQ_{\mu\nu}$, $\hat\CQ_\alpha{}^{\mu\nu}$ and $\hat\CT^\alpha{}_{\mu\nu}$ such that $\hat Q_{\mu\nu} - \hat\CQ_{\mu\nu}$, $\hat Q_\alpha{}^{\mu\nu} - \hat \CQ_\alpha{}^{\mu\nu}$ and $T^\alpha{}_{\mu\nu} - \hat \CT^\alpha{}_{\mu\nu}$ depend linearly on $\omega_{\mu\nu}$ and for which  
\begin{align}
    \tau_\nu \hat\CQ_\alpha{}^{\mu\nu} = - \hat\CQ_\alpha{}^\mu, \quad \tau_\alpha \hat\CT^\alpha{}_{\mu\nu} = -2\hat\CQ_{[\mu\nu]}.\label{eq_reduced_identities_Gal}
\end{align}
Given a choice of $(v^\mu, \gamma_{\mu\nu})$, we introduce $\tilde\omega_{\mu\nu} \coloneqq h^\sigma{}_{\mu} h^\lambda{}_\nu \omega_{\sigma\lambda}$ and $A_\mu \coloneqq \omega_{\mu\nu} v^\nu$. The linear dependency of the reduced quantities on $\omega_{\mu\nu}$ implies the general form
\begin{align}
    \hat Q_{\mu\nu}
        &= \hat\CQ_{\mu\nu} + a_1\, \tilde\omega_{\mu\nu} + a_2\, \tau_\mu A_\nu + a_3\, \tau_\nu A_\mu\,,\\
    \hat Q_\alpha{}^{\mu\nu}
        &= \hat\CQ_\alpha{}^{\mu\nu} + b_1\, v^{(\mu}\tilde\omega^{\nu)}{}_\alpha + b_2\, v^{(\mu} A^{\nu)} \tau_\alpha + b_3\, v^\mu v^\nu A_\alpha + b_4\,  A^{(\mu}h^{\nu)}{}_\alpha\,,\\
    T^\alpha{}_{\mu\nu}
        &= \hat\CT^\alpha{}_{\mu\nu} + c_1 \, v^\alpha \tilde\omega_{\mu\nu} + c_2 \,\tau_{[\mu} \tilde\omega_{\nu]}{}^\alpha + c_3\, v^\alpha \tau_{[\mu} A_{\nu]} + c_4\, h^\alpha{}_{[\mu} A_{\nu]}\,.
\end{align}
The general form of an affine connection becomes
\begin{equation}
\begin{aligned}\label{eq_gene_Gamma_abc}
    \Gamma^\alpha_{\mu\nu} - {^{v,\tau}}\Gamma^\alpha_{\mu\nu} &=h^\alpha{}_\beta \hat\CQ_{(\mu\nu)}{}^\beta - \frac{1}{2} \hat\CQ^\alpha{}_{\mu\nu} - v^\alpha \hat\CQ_{(\mu\nu)} - \hat\CT_{(\mu\nu)}{}^\alpha + \frac{1}{2} \hat\CT^\alpha{}_{\mu\nu} + 2\tau_{(\mu} \kappa_{\nu)\beta}h^{\alpha\beta} \\
    &\qquad + v^\alpha\left[-(a_2 + a_3) \tau_{(\mu} A_{\nu)} + \frac{c_3}{2} \tau_{[\mu} A_{\nu]} + \frac{c_1}{2} \tilde \omega_{\mu\nu}\right] + 
    \frac{1}{2}\left(b_4-c_4\right)\,\gamma_{\mu\nu} A^\alpha \\
    &\qquad+ \frac{1}{2}\left(b_4 + c_4\right)\, h^\alpha{}_{(\mu} A_{\nu)}
    + \frac{c_4}{2}h^\alpha{}_{[\mu} A_{\nu]}+ \frac{c_2}{2} \tau_\mu \tilde\omega_\nu{}^\alpha\,.
\end{aligned}
\end{equation}
Using the constraints~\eqref{eq_reduced_identities_Gal} and the identities~\eqref{eq_identities_Gal}, we get
\begin{align}
    b_1 = 2a_1, && b_2 = -2a_2, && b_3 = 0, && c_1 = 2(1-a_1), && c_3 = 2\left(a_3 - a_2 - 2\right),
\end{align}
with $b_4$, $c_2$ and $c_4$ being unconstrained. We see that there is a 6-parameter freedom in defining the reduced quantities such that~\eqref{eq_reduced_identities_Gal} hold. In general it is not possible to choose the free parameters such that all the reduced quantities become gauge independent. The only way to reduce the degrees of freedom is to assume either metricity or torsion-freeness:
\begin{itemize}
    \item Assuming full metricity with $\hat Q_{\mu\nu} = \hat\CQ_{\mu\nu} =0$ and $\hat Q_\alpha{}^{\mu\nu} = \hat\CQ_\alpha{}^{\mu\nu} = 0$, we get $c_1 = 2$ and $c_3=-4$ and we get
\begin{align}
    T^\alpha{}_{\mu\nu}
    &= 
    \hat\CT^\alpha{}_{\mu\nu} +2 \, v^\alpha \omega_{\mu\nu} + c_2 \,\tau_{[\mu} \tilde\omega_{\nu]}{}^\alpha + c_4\, h^\alpha{}_{[\mu} A_{\nu]}; \, (c_2\ \ \text{and}\ c_4\ \ \text{are free}),
\\
\begin{split}
    \Gamma^\alpha_{\mu\nu} - {^{v,\tau}}\Gamma^\alpha_{\mu\nu}
        &= - \hat\CT_{(\mu\nu)}{}^\alpha + \frac{1}{2} \hat\CT^\alpha{}_{\mu\nu} + 2\tau_{(\mu} \kappa_{\nu)\beta}h^{\alpha\beta} + v^\alpha\,  \omega_{\mu\nu} \\
        &\quad  + \frac{c_4}{2}\left(-\gamma_{\mu\nu} A^\alpha + h^\alpha{}_\mu A_\nu\right) + \frac{c_2}{2} \tau_\mu \tilde\omega_\nu{}^\alpha
\,.
\end{split}
\end{align}
    \item Assuming no torsion with $T^\alpha{}_{\mu\nu} = \hat\CT^\alpha{}_{\mu\nu} = 0$, we get
\begin{align}
     \hat Q_{\mu\nu}
        &= \hat\CQ_{\mu\nu} + \tilde\omega_{\mu\nu} + a_2\, \tau_\mu A_\nu + (2+a_2)\, \tau_\nu A_\mu \,; \quad  (a_2\ \ \text{is free}),        
\\
    \hat Q_\alpha{}^{\mu\nu}
        &= \hat\CQ_\alpha{}^{\mu\nu} + 2 v^{(\mu}\tilde\omega^{\nu)}{}_\alpha - 2a_2\, v^{(\mu} A^{\nu)} \tau_\alpha + b_4\,  A^{(\mu}h^{\nu)}{}_\alpha \,; \quad (b_4\ \ \text{is free}),        
\\
\begin{split}
    \Gamma^\alpha_{\mu\nu} - {^{v,\tau}}\Gamma^\alpha_{\mu\nu} 
    &=
    h^\alpha{}_\beta \hat\CQ_{(\mu\nu)}{}^\beta - \frac{1}{2} \hat\CQ^\alpha{}_{\mu\nu} - v^\alpha \hat\CQ_{(\mu\nu)} 
    + 2\tau_{(\mu} \kappa_{\nu)\beta}h^{\alpha\beta} 
\\
&\quad
    - 2(1 + a_2 ) \, v^\alpha \tau_{(\mu} A_{\nu)}
\,.
\end{split}
\end{align}
The reduced quantities obtained from $\tayl{0}{\Gamma}^{\,\alpha}_{\mu\nu}$ by imposing $\hat\CQ_{\mu\nu} = 0 = \hat\CQ_\alpha{}^{\mu\nu}$ (taking $\T\tau\wedge\T\dd\T\tau = 0$ from global hyperbolicity) in the non-relativistic limit correspond to $a_2 = -2$ and $b_4 = 0$.
\end{itemize}

\subsection{Carrollian case}\label{app_reduced_quantities_Car}

The goal is to define the reduced quantities $\check \CQ_{\mu}{}^\nu$, $\check \CQ_{\alpha\mu\nu}$ and $\check \CT^\alpha{}_{\mu\nu}$ such that $\check Q_{\mu}{}^{\nu} - \check \CQ_{\mu}{}^{\nu}$, $\check Q_{\alpha\mu\nu} - \check \CQ_{\alpha\mu\nu}$ and $T^\alpha{}_{\mu\nu} - \check \CT^\alpha{}_{\mu\nu}$ depend linearly on $\Theta_{\mu\nu}$ and for which  
\begin{align}\label{eq_reduced_identities_Car}
    v^\nu \check \CQ_{\alpha\mu\nu} = - \check \CQ_{\alpha\mu}, \quad v^\alpha \check \CT_{(\mu\nu)\alpha} = \frac{1}{2} v^\alpha \check \CQ_{\alpha\mu\nu} + \check \CQ_{(\mu\nu)}.
\end{align}
We introduce the trace $\theta \coloneqq \Theta_{\mu\nu} h^{\mu\nu}$ and the traceless part $A_{\mu\nu} \coloneqq \Theta_{\mu\nu} - \frac{\theta}{3} \gamma_{\mu\nu}$ of $\Theta_{\mu\nu}$. Both $\theta$ and $A_{\mu\nu}$ are gauge independent quantities (contrary to $A_\mu$ in the Galilean case). The linear dependency of the reduced quantities on $\Theta_{\mu\nu}$ implies the general form
\begin{align}
    \check Q_{\mu}{}^{\nu}
        &= \check \CQ_{\mu}{}^{\nu} + a_1 \, A_\mu{}^\nu + \left(a_2\, \tau_\mu v^\nu + a_3\, h^\nu{}_\mu\right) \theta,\\
    \check Q_{\alpha\mu\nu}
        &= \check \CQ_{\alpha\mu\nu} + b_1\, \tau_\alpha A_{\mu\nu} + b_2\, \tau_{(\mu}A_{\nu)\alpha} + \left[b_3\, \tau_\mu \tau_\nu \tau_\alpha + b_4\, \gamma_{\mu\nu} \tau_\alpha + b_5\, \tau_{(\mu}\gamma_{\nu)\alpha}\right]\theta,\\
    T^\alpha{}_{\mu\nu}
        &= \check \CT^\alpha{}_{\mu\nu} + c_1\, \tau_{[\mu}A_{\nu]}{}^\alpha + c_2\, \tau_{[\mu} h_{\nu]}{}^\alpha \theta.
\end{align}
The general form of an affine connection becomes
\begin{equation}
\begin{aligned}
    \Gamma^\alpha_{\mu\nu} - {^{v,\tau}}\Gamma^\alpha_{\mu\nu}
        &=
        -\check \CQ_{(\mu\nu)}{}^\alpha + \frac{1}{2} \check \CQ^\alpha{}_{\mu\nu} - \check \CT_{(\mu\nu)}{}^\alpha + \frac{1}{2} \check \CT^\alpha{}_{\mu\nu} - v^\alpha \chi_{\mu\nu}
\\
&\quad 
        - \left( b_1 + \frac{c_1}{2}\right) \tau_{(\mu} A_{\nu)}{}^\alpha + \frac{c_1}{2} \tau_{[\mu}A_{\nu]}{}^\alpha
\\
&\quad
        -
        \left[
            \left(b_4 + \frac{c_2}{2}\right) \tau_{(\mu} h_{\nu)}{}^\alpha - \frac{c_2}{2} \tau_{[\mu}h_{\nu]}{}^\alpha
        \right] \theta.
\end{aligned}
\end{equation}

Using the constraints~\eqref{eq_reduced_identities_Car} and the identities~\eqref{eq_identities_Car}, we get
\begin{align}
    b_2 = -2a_1, && b_5 = -2a_3, && b_3 = 0, && c_1 = 2(1-a_1) - b_1, && c_2 = \frac{2}{3} - b_4 - 2a_3.
\end{align}
We see that there is a 5-parameter freedom in defining the reduced quantities such that~\eqref{eq_reduced_identities_Car} hold. As in the Galilean case, in general it is not possible to choose the free parameters such that all the reduced quantities become gauge independent. The only way to reduce the degrees of freedom is to assume either metricity or torsion-freeness:

\begin{itemize}
    \item Assuming metricity with $\check Q_{\mu}{}^\nu = \check \CQ_{\mu}{}^\nu =0$ and $\check Q_{\alpha\mu\nu} = \check \CQ_{\alpha\mu\nu} = 0$, we have $c_1 = 2$ and $c_2 = 2/3$, and we get
\begin{align}
    T^\alpha{}_{\mu\nu}
        &= \check \CT^\alpha{}_{\mu\nu} + \, 2\tau_{[\mu}\Theta_{\nu]}{}^\alpha\,; \quad (\text{no degrees of freedom left}), \label{eq_reduced_T_Car_app}\\
    \Gamma^\alpha_{\mu\nu} - {^{v,\tau}}\Gamma^\alpha_{\mu\nu}
    &= - \check \CT_{(\mu\nu)}{}^\alpha + \frac{1}{2} \check \CT^\alpha{}_{\mu\nu} - v^\alpha \chi_{\mu\nu} - \tau_{\nu} \Theta_{\mu}{}^\alpha.
\end{align}
In this case, there is no freedom remaining. So reduced torsion for a metric Carrollian connection is uniquely defined.
    \item Assuming no torsion with $T^\alpha{}_{\mu\nu} = \check \CT^\alpha{}_{\mu\nu} = 0$, we get
\end{itemize}
\begin{align}
    \check Q_{\mu}{}^{\nu}
        &= \check \CQ_{\mu}{}^{\nu} + a_1 \, A_\mu{}^\nu + \left(a_2\, \tau_\mu v^\nu + a_3\, h^\nu{}_\mu\right) \theta\,; \ \ (a_1, \ a_2, \ \text{and}\ a_3\ \text{are free}),
\\
\begin{split}
    \check Q_{\alpha\mu\nu}
        &= \check \CQ_{\alpha\mu\nu} + 2(1-a_1)\, \tau_\alpha A_{\mu\nu} \\
        &\quad - 2a_1\, \tau_{(\mu}A_{\nu)\alpha} + \left[(\frac{2}{3}-2a_3)\, \gamma_{\mu\nu} \tau_\alpha -2a_3\, \tau_{(\mu}\gamma_{\nu)\alpha}\right]\theta,
\end{split}
\\
\begin{split}
    \Gamma^\alpha_{\mu\nu} - {^{v,\tau}}\Gamma^\alpha_{\mu\nu}
    &=-\check \CQ_{(\mu\nu)}{}^\alpha + \frac{1}{2} \check \CQ^\alpha{}_{\mu\nu} - v^\alpha \chi_{\mu\nu} 
\\
&\quad
    -2(1-a_1)\,\tau_{(\mu} A_{\nu)}{}^\alpha
        -(\frac{2}{3} - 2a_3)\, \tau_{(\mu} h_{\nu)}{}^\alpha \, \theta
\,.
\end{split}
\end{align}
\begin{itemize}
    \item[] The reduced quantities obtained from $\tayl{0}{\Gamma}^{\,\alpha}_{\mu\nu}$ by imposing $\check \CQ_\mu{}^\nu = 0 = \check \CQ_{\alpha\mu\nu}$ in the Carrollian limit correspond to $a_1 = 1$, $a_2 = 0$ and $a_3 = \frac{1}{3}$.
\end{itemize}

\bibliographystyle{apsrev4-1}
{\small \bibliography{refs}}

\end{document}